\newcommand{\be}{\begin{equation}}
\newcommand{\ee}{\end{equation}}
\newcommand{\bea}{\begin{eqnarray}}
\newcommand{\eea}{\end{eqnarray}}

\def\ov{\over  }

\documentclass[a4paper]{article} 

\usepackage{epsfig}
\usepackage{graphicx}
\usepackage{epsfig,cite}

\input epsf
\usepackage{amsmath,amsfonts,epsfig,color,latexsym}
\usepackage{amssymb}
\usepackage{amsfonts}
\usepackage{amssymb}

\input epsf
\usepackage{amsmath,amsfonts,epsfig,color,latexsym}
\usepackage{amssymb}
\usepackage{amsfonts}
\usepackage{amssymb}
\begin{document}

\centerline{\LARGE{\bf{Sommerfeld enhancement for a Yukawa potential}}}

\vskip 0.3cm
\centerline{\bf Roberto Iengo \footnote{iengo@sissa.it}}

\vskip .3cm
\centerline{\it  International School for Advanced Studies (SISSA)}
\centerline{\it Via Beirut 2-4, I-34013 Trieste, Italy} 
\centerline{\it  INFN, Sezione di Trieste}
\vskip0.3cm

{\bf Abstract.} We show how easy it is to get the Sommerfeld enhancement for a Yukawa potential, for definite partial waves,
beyond  the S wave analyzed in previous literature. In particular,
we report results for the P wave (for which there is a resonant pattern
and the enhancement can be of several orders of magnitude even far from resonances) 
that could be relevant for the analysis of experimental cosmic rays data
possibly signaling the annihilation  of dark matter particles.

\section{The general formula}

Here we derive the formula for the Sommerfeld enhancement for a definite partial wave $l$
in the case of an attractive  Yukawa potential $-{\alpha e^{-\mu r}\ov r}$.
\vskip0.2cm
 This is the non-relativistic limit of a process in which two particles attract each other by exchanging repeatedly  a  massive vector (or also scalar) boson,
 before undergoing some annihilation reaction. The Sommerfeld enhancement could be relevant for the interpretation of cosmic rays data as due to the annihilation of dark matter particles, 
and it has been invoked in various papers \cite{Hisa},\cite{ Stru},\cite{Stru2},\cite{Mar}, \cite{ArkHam},\cite{Lat},\cite{Cli},\cite{Rob}. 
However, as far as we know, it has not been studied for partial waves higher than $l=0$, in particular 
for the $l=1$ P partial wave, which could be particularly relevant.    
 
 In general one gets a matrix valued potential. In the analysis below we assume that 
that  the matrix can be diagonalized and  that it has a Yukawa form for the eigenstates of the  two particles. We consider therefore 
a definite eigenstate.

The starting point is eq.(2.35) of ref\cite{ie}:
\be
A_l(p,p')=\sqrt{\pi\ov 2}{1\cdot 3\cdots (2l+1)\ov l!}{1\ov p}
({d\ov dr})^l R_{p,l}(r)|_{r=0} a_{0,l}(p')
\label{start}
\ee

where $A_l(p,p')$ is the $l$ partial wave amplitude for the process described above,
$p$ being the CM momentum of the interacting particles and $p'$ the one of the reaction's
products;  \\
$A_{0,l}(p,p')\equiv p^l a_{0,l}(p')$ is the $bare$ (i.e. for $\alpha=0$) 
partial wave reaction amplitude. 

We have assumed that the bare reaction amplitude 
is proportional to $p^l$, which, for small $p$, is the typical dependence  on the incoming
momentum of the $l$ partial wave.

$R_{p,l}(r)$ is the solution of the partial wave Schroedinger equation:
\be
-{1\ov 2m_r}({d^2 R_{p,l}\ov dr^2}+{2\ov r}{dR_{p,l}\ov dr}-{l(l+1)R_{p,l}\ov r^2})-
({p^2\ov 2m_r}+{\alpha e^{-\mu r}\ov r})R_{p,l}=0
\ee
normalized such that 
\be
\int_0^\infty r^2 dr R_{q,l}(r)R_{p,l}(r)=\delta (p-q)
\ee
Here $m_r=m/2$ is the reduced mass and $m$ is the mass of the two particles.

We know from textbooks (see for instance \cite{LL}) that this normalization corresponds to
the asymptotic behavior
\be
{R_{p,l}(r)}_{r\to\infty} \to \sqrt{2\ov\pi}{\sin(pr-{l\pi\ov 2}+\delta_l)\ov r}
\label{normR}
\ee

Let us define $x=pr$ and put $R_{p,l}(r)=N px^l\Phi_l(x)$; the equation for $\Phi_l$ is:
\be
\Phi_l''+{2(l+1)\ov x}\Phi_l'+({2 a e^{-bx}\ov x}+1)\Phi_l=0
\label{eqphi}
\ee
where $a\equiv \alpha /v$, $b\equiv \mu /(m_r v)$ and $v=p/m_r$ is the relative velocity. 

Suppose we solve this equation with the initial conditions 
\be
\Phi_l(0)=1 ~~~~~ \Phi_l'(0)=-a/(l+1)
\label{init}
\ee
(the condition for $\Phi_l'(0)$ is dictated by the equation
for a regular solution). Then the asymptotic behavior will be
\be
{x^{l+1}\Phi_l(x)}_{x\to\infty}\to C\sin(x-{l\pi\ov 2}+\delta_l)
\label{asy}
\ee
In order to agree with the normalization of eq.(\ref{normR}) we have to put 
$N=\sqrt{2\ov\pi}{1\ov C}$. Substituting in eq.(\ref{start}) we get
\be
A_l(p,p')={1\cdot 3\cdots (2l+1)\ov C} p^la_{0,l}={1\cdot 3\cdots (2l+1)\ov C} A_{0,l}(p,p')
\ee
In conclusion, by defining the Sommerfeld enhancement $enh_l$ for the $l$ partial wave cross-section (or equivalently for the rate) as
\be
\sigma_l=enh_l\cdot\sigma_{0,l}
\ee
 we get
\be
enh_l=({1\cdot 3\cdots (2l+1)\ov C})^2
\label{result}
\ee
where $C$ is obtained by looking at the asymptotic behavior eq.(\ref{asy}) of the solution of
eq.(\ref{eqphi}) with the initial conditions eq.(\ref{init}). 
$enh_l$ depends on the two parameters $a$ and $b$. It is not necessary to determine $\delta_l$.

Another equivalent  strategy is to put $R_{p,l}(r)=N p\varphi_l(x)/x$; the equation for $\varphi_l$ is:
\be
\varphi_l''+(1+{2 a\ov x}e^{-bx}-{l(l+1)\ov x^2})\varphi_l=0
\label{eqb}
\ee
If one solves this equation with the initial conditions corresponding to 
\be
{\varphi_l(x)}_{x\to 0}\to x^{l+1}
\label{initb}
\ee 
then the 
asymptotic behavior will be
\be
{\varphi_l(x)}_{x\to\infty}\to C\sin(x-{l\pi\ov 2}+\delta_l)
\label{asyb}
\ee
with the same $C$ of eq.(\ref{asy}) giving the enhancement as in eq.(\ref{result}).

\section{Computations for $l=1$}

In principle it is easy to get $C$: for instance one can use the NDSolve instruction of 
Mathematica to get the numerical solution of eq.(\ref{eqphi}) with the initial 
conditions eq.(\ref{init}), or equivalently of eq.(\ref{eqb})  with initial conditions eq.(\ref{initb}).
In order to find $C$ one takes $F_l(x)\equiv x^{l+1}\Phi_l(x)$ or else $F(x)\equiv \varphi(x)$,
and  one
plots $F_l(x)^2+F_l(x-\pi/2)^2$ for large $x$: when this is constant it is equal to $C^2$. 
We follow the strategy of eqs.(\ref{eqb},\ref{initb}), which provides more clean numerical results.

It is expected that this procedure should work less well for $b$ very low and $a$ very large because
in this case the asymptotia is reached for very large $x$ and the numerical solution
accumulates errors. However, for $b=0$ and for any $a$ we already have  the exact result derived analytically 
in ref. \cite{ie}:
\be
enh_l=\prod_{s=1}^l(s^2+a^2) e^{\pi a}{\pi a\ov sinh(\pi a) l!^2}
\ee

In practice, this works well for $l=0$, see for instance Fig. 1.

\begin{figure}[ht]\label{fig.1}
\centerline{\epsfig{file=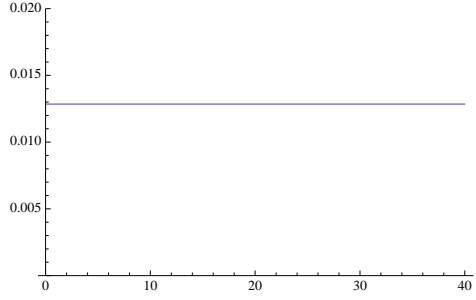,width=.4\textwidth}}
\caption{$F_0(x)^2+F_0(x-\pi/2)^2$ for $\alpha=10^{-2},~\mu=1Gev,~m=10^3Gev,~v=2\times 10^5$.}
\end{figure}

\begin{figure}[ht]\label{fig.2}
\centerline{\epsfig{file=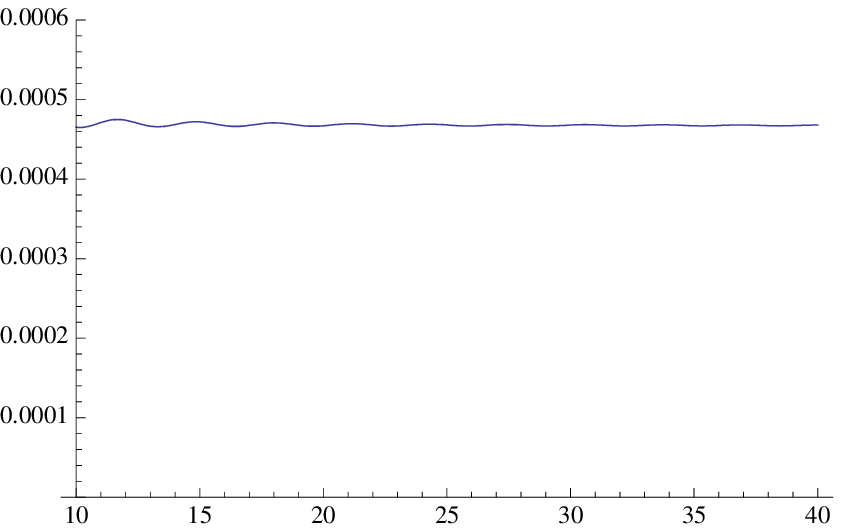,width=.4\textwidth}\ \epsfig{file=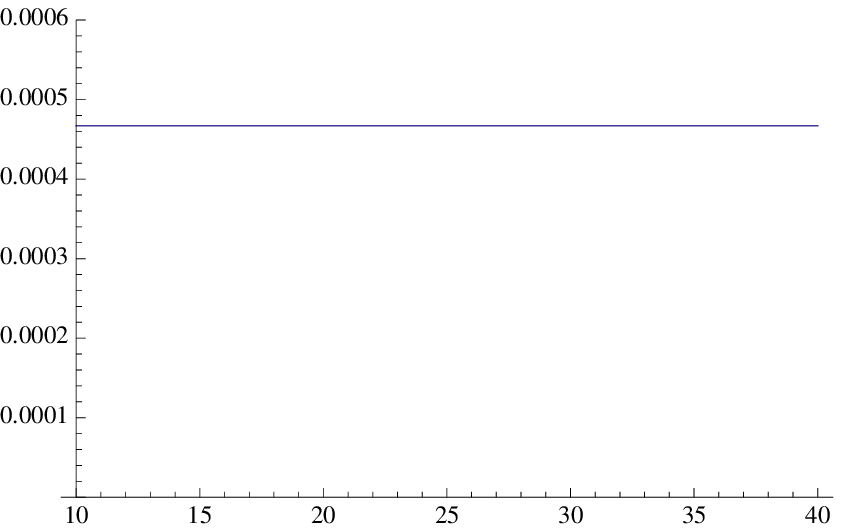,width=.4\textwidth}}
\caption{ $F_1(x)^2+F_1(x-\pi/2)^2$ (left) and $j(x)$ (right) for the same parameters of Fig.1.}
\end{figure}

\begin{figure}[ht]\label{fig.3}
\centerline{\epsfig{file=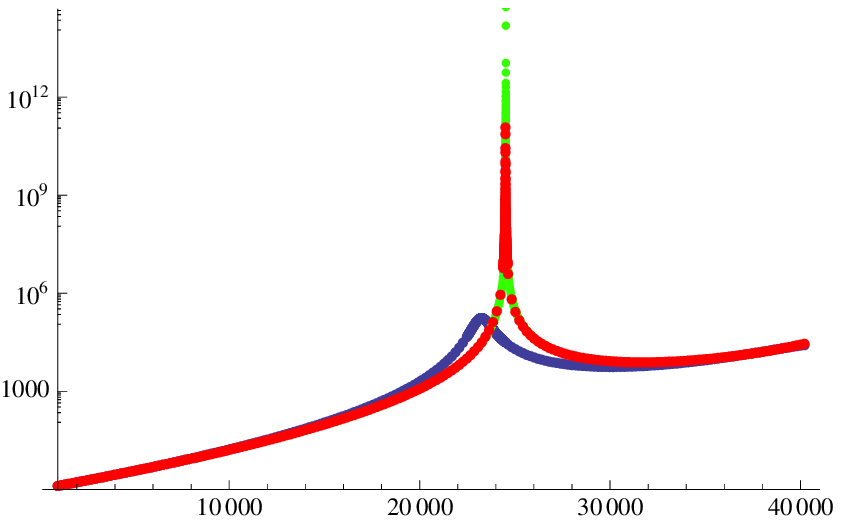,width=.6\textwidth}}
\caption{ $enh_1$ as a function of $m(Gev)$ for $v_{single~particle}=10^{-3}$ blue, $10^{-4}$ red ,
$10^{-5}$ green. Here $\alpha=1/30$, $\mu=90 Gev$.}
\centerline{\epsfig{file=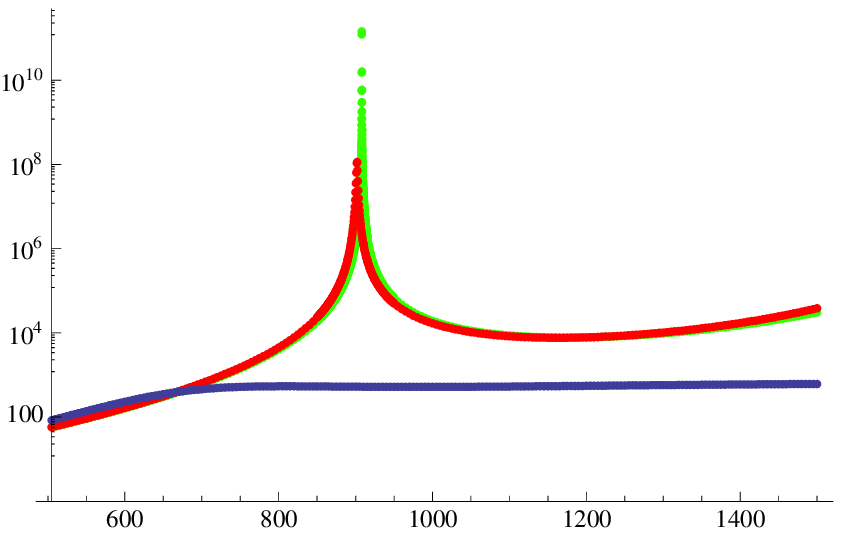,width=.6\textwidth}}
\caption{ $enh_1$ as a function of $m(Gev)$ for $v_{single~particle}=10^{-3}$ blue, $10^{-4}$ red ,
$10^{-5}$ green. Here $\alpha=1/100$, $\mu=1 Gev$.}
\end{figure}

For $l=1$ the quantity $F_l(x)^2+F_l(x-\pi/2)^2$ may sometimes continue to show
decreasing oscillations: here it is convenient to take into account the sub-leading term in the asymptotic
expansion (\ref{asyb}), which we know to be the free wave function up to the phase shift $\delta_l$.
Therefore, the improved version of (\ref{asyb}) for $l=1$ is:
\be
{\varphi_1(x)}_{x\to\infty}\to C\cdot \big(\sin(x-{l\pi\ov 2}+\delta_l)+{\cos(x-{l\pi\ov 2}+\delta_l)\ov x}\big)
\ee
To get $C^2$, one defines, in three steps, with $F_1(x):= \varphi_1(x)$,  
\bea
k(x) &:=& {(\pi^2-16 x^2)^2\ov(8\pi(\pi^2-16x^2))}(F_1(x+\pi/4)^2+F_1(x-\pi/4)^2), \nonumber \\ 
h(x) &:=& (\pi^3-4\pi x^2)(k(x+\pi/4)+k(x-\pi/4), \nonumber \\
j(x) &:=&-8{h(x+\pi/4)+h(x-\pi/4)\ov  (\pi^2-16x^2)(3\pi^2+16(1+x^2))}. \nonumber
\eea
For $x$ large (say $x>30$), $j(x)$ quickly converges to a constant  which  equals $C^2$.  
(For $l>1$ the free wave function is more complicated and one should do more steps).
It must be said that the results of the improved procedure differ little
from what could be obtained simply by finding by eye the average of the 
oscillations of  $F_l(x)^2+F_l(x-\pi/2)^2$, see Fig. 2.

The case of the S $l=0$ wave has been discussed at length in the literature, and it has been found 
a resonant pattern, see in particular refs. \cite{Lat} and \cite{Rob}. We have verified that we get the same results.

Here we present some numerical result for the P $l=1$ wave, which also shows a resonant pattern.

In Fig.3 we show the enhancement eq.(\ref{result}) for $l=1$, taking the values of the parameters 
used in the numerical evaluations for the S wave in ref. \cite{Rob} for the range of ref. \cite{Lat}, 
that is $\alpha=1/30,~\mu=90Gev$, as a function of $m$ (expressed in $Gev$)
for $v_{single~particle}=10^{-3},10^{-4},10^{-5}$.

In Fig.4 we show the enhancement eq.(\ref{result}) for $l=1$, taking the values of the parameters 
used in the numerical evaluations for the S wave in ref. \cite{Rob} for the range of ref. \cite{ArkHam}, 
that is $\alpha=1/100,~\mu=1Gev$, as a function of $m$ (expressed in $Gev$)
for $v_{single~particle}=10^{-3},10^{-4},10^{-5}$.


\end{document}